\documentclass[aip,rsi,amsmath,reprint,amssymb,groupedaddress]{revtex4-2}

\pdfoutput=1

\usepackage{graphicx}
\usepackage{textcomp}

\usepackage[normalem]{ulem}
\usepackage{color}
\newcommand{\MS}[1]{\textcolor{black}{#1}}

\begin{document}

\title{Characterization of harmonic modes and parasitic resonances in multi-mode superconducting coplanar resonators}

\author{Cenk~Beydeda}
\email[]{cenk.beydeda@pi1.uni-stuttgart.de}
\author{Konstantin~Nikolaou}
\author{Marius~Tochtermann}
\author{Nikolaj~G.~Ebensperger}
\author{Gabriele~Untereiner}
\author{Ahmed~Farag}
\author{Philipp~Karl}
\author{Monika~Ubl}
\author{Harald~Giessen}
\author{Martin~Dressel}
\author{Marc~Scheffler}
\email[]{marc.scheffler@pi1.physik.uni-stuttgart.de}
\affiliation{Physikalisches Institut, Universit\"at Stuttgart, 70569 Stuttgart, Germany}

\date{28.03.2023}

\begin{abstract}
Planar superconducting microwave transmission line resonators can be operated at multiple harmonic resonance frequencies. This allows covering wide spectral regimes with high sensitivity, as it is desired e.g.\ for cryogenic microwave spectroscopy. 
A common complication of such experiments is the presence of undesired \lq spurious\rq{} additional resonances, which are due to standing waves within the resonator substrate or housing box.
Identifying the nature of individual resonances (\lq designed\rq{} vs.\ \lq spurious\rq ) can become challenging for higher frequencies or if elements with unknown material properties are included, as is common for microwave spectroscopy.
Here we discuss various experimental strategies to distinguish designed and spurious modes in coplanar superconducting resonators that are operated in a broad frequency range up to 20~GHz. These strategies include tracking resonance evolution as a function of temperature, magnetic field, and microwave power. We also demonstrate that local modification of the resonator, by applying minute amounts of dielectric or ESR-active materials, lead to characteristic signatures in the various resonance modes, depending on the local strength of the electric or magnetic microwave fields.

\end{abstract}

\maketitle


\section{Introduction}

Planar superconducting resonators, fabricated from superconducting thin films on insulating substrates, play an important role for cryogenic on-chip applications and in various research fields. In quantum information processing, superconducting resonators couple microwave photons to individual solid-state quantum bits or ensembles of quantum systems.\cite{Wallraff2004,Sillanpaeae2007,Majer2007,Schuster2010,Kubo2010,Huebl2013,Ghirri2015,Gu2017,Hattermann2017}
In astronomy and particle physics, highly sensitive kinetic inductance detectors (KIDs) can easily be multiplexed.\cite{Day2003,Zmuidzinas2012,Battistelli2015,Adam2018}
In solid state spectroscopy,\cite{Scheffler2013,Hafner2014,McRae2020} planar superconducting resonators probe the microwave properties of numerous material classes of interest, ranging from conventional\cite{DiIorio1988,Oates1991,Andreone1993,Zemlicka2015,Driessen2012,Beutel2016,Thiemann2018a,Thiemann2018b,Manca2019} and unconventional superconductors \cite{Anlage1989,Langley1991,Revenaz1994,Porch1995,Zaitsev2001,Wang2007,Ghigo2012,Scheffler2015,Ghigo2016} to heavy-fermion metals,\cite{Scheffler2013,Parkkinen2015} quantum paraelectrics,\cite{Davidovikj2017,Engl2019} various magnetic and spin systems,\cite{Wallace1998,Bushev2011,Malissa2013,Bondorf2018,Golovchanskiy2018,Ranjan2020,Miksch2021} and dielectric thin films.\cite{OConnell2008,Wisbey2019,Ebensperger2019}

Realization of on-chip superconducting resonators can follow different approaches, such as lumped element resonators\cite{Doyle2008,FornDiaz2010} or transmission line resonators.\cite{Frunzio2005,Goeppl2008} The latter employs one of various transmission line geometries (e.g.\ coplanar, microstrip, or stripline); here a line segment of a certain length with open or short ends defines a one-dimensional resonator.
The higher resonance modes of transmission line resonators are harmonics, which in the simplest case are spaced equally in frequency, and they have transverse field distributions corresponding to the fundamental mode. These properties are advantageous for microwave spectroscopy applications, because they allow to conveniently cover a rather wide frequency range combined with high sensitivity and straightforward data analysis.\cite{DiIorio1988,Andreone1997,Scheffler2013,Hafner2014,Zou2017} 
Typical spectral ranges span from 1 to 20~GHz and beyond.\cite{Wang2007,Davidovikj2017,Rausch2018,Thiemann2018b,Manca2019}
If one operates a superconducting on-chip resonator in such a broad frequency range, one typically encounters various additional resonances that are undesired and that stem e.g.\ from standing waves in the dielectric substrate or in the metallic sample holder box, or from 
asymmetric slotline modes.\cite{Schuster2010,Wenner2011,McRae2020}
For a superconducting microwave device operating at a single frequency or in a narrow frequency range,\cite{Adam2018} the microwave environment (e.g.\ sample box) can often be optimized such that all the parasitic modes are shifted to frequency ranges that are not relevant for the particular device, usually this means to higher frequencies.\cite{Wenner2011,McConkey2018} Also slotline modes can often be eliminated by e.g.\ bridging wirebonds.
But for spectroscopy studies, avoiding such parasitic resonances completely usually is not possible. 
Then it is crucial to identify which of the detected resonances are the designed resonator harmonics and which are the parasitic modes. This is straightforward if the harmonics are evenly distributed in frequency. 
But the material properties to be determined in microwave spectroscopy can exhibit substantial frequency dependence,\cite{Viana1994,Sluchanko2000,Tran2002,Turner2003,Scheffler2005c,Engl2019} and thus the resulting resonator frequencies are not known beforehand and might not be spaced evenly in frequency. 
In such cases, identifying whether an observed resonance is one of the designed modes or parasitic, can become challenging.
Here we present various strategies how one can characterize such higher-frequency modes and determine their nature.

\section{Experiment and Data Analysis}

\begin{figure}
	\centering
	\includegraphics[width=\linewidth]{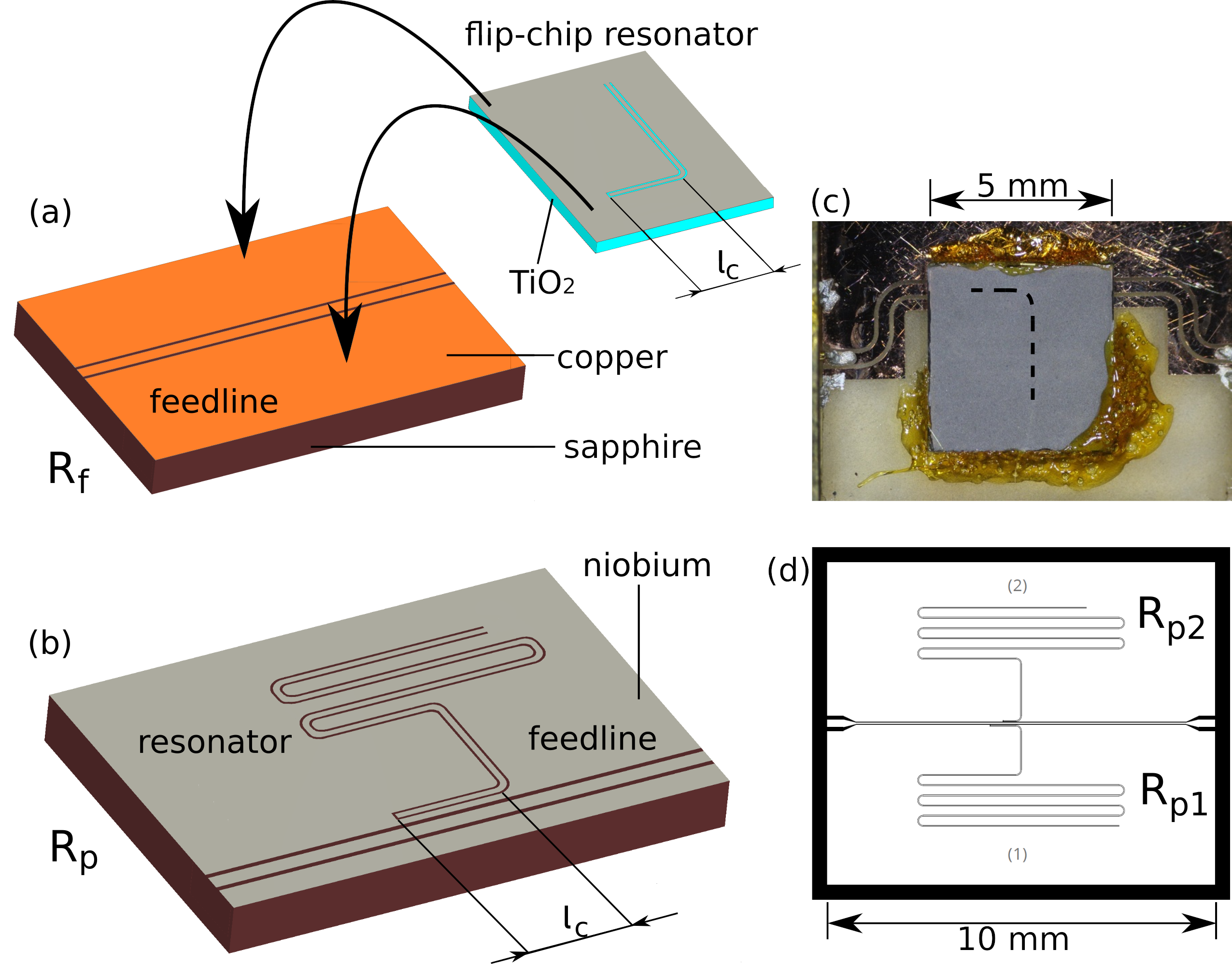}
	\caption{Schematic resonator designs of the (a) flip-chip setup (labeled R$_\textrm{f}$) and the (b) on-plane setup (labeled R$_\textrm{p}$) and (c) photograph and (d) design of the actual devices. For R$_\textrm{f}$ the resonator chip (Nb on TiO$_2$) is held face-to-face slightly above a separate feedline chip (Cu on sapphire), while for R$_\textrm{p}$ two resonators and a feedline are fabricated from the same Nb layer on a sapphire substrate. The dashed line in (c) indicates the position of the resonator on the lower side of the TiO$_2$ substrate. The two resonators in (d) are labeled R$_\text{p1}$ and R$_\text{p2}$ for distinction. $l_c$ is the coupling length of the resonators.}
	\label{fig:Setup}
\end{figure}

We present and discuss data that are mostly obtained with two different coplanar waveguide (CPW) resonator designs, labeled R$_\textrm{f}$ and R$_\textrm{p}$, as shown in Fig.\ \ref{fig:Setup}. Each resonator is fabricated by optical lithography from a Nb layer with 300\,nm thickness on a dielectric substrate. We employ $\lambda/4$-resonators (with $\lambda$ the wavelength in the CPW) that are coupled to their feedlines by parallel straight sections of the CPWs of length $l_c$.

The first case, R$_\textrm{f}$ shown in Figs.\ \ref{fig:Setup}(a) and (c), consists of a coplanar resonator fabricated on a TiO$_2$ substrate, forming the so-called flip-chip, which is mounted above a copper feedline that is deposited on a separate sapphire (Al$_2$O$_3$) chip. The distance between both chips is $\approx$~50~$\mu$m, and the coupling arm of the resonator on the flip-chip and the copper feedline face each other.\cite{Wendel2020} The dielectric constant of TiO$_2$, between 110 and 260 depending on crystallographic direction,\cite{Klein1995,Zuccaro1997,Tobar1998} is rather high, and thus harmonic and parasitic modes incorporating the TiO$_2$ can occur at comparably low frequencies. 

The second case, R$_\textrm{p}$ shown in Figs.\ \ref{fig:Setup}(b) and (d), employs a sapphire chip with feedline and two resonators respectively labeled R$_\text{p1}$ and R$_\text{p2}$ arranged in the same plane, i.e.\ both resonators can be addressed with a single microwave line like multiplexed devices.\cite{Day2003,Adam2018} 
These CPW resonators have meander shape to allow low fundamental frequencies for a small chip area, which in fact is a strategy to suppress parasitic box modes.


The $\lambda/4$-resonators of both chips, R$_\textrm{f}$ and R$_\textrm{p}$, support resonances at odd multiples $n=1,3,5,...$ of the fundamental mode frequency $f_0$:
\begin{equation}
f_\textrm{n} = n f_0 = n\cdot c(4l\sqrt{\epsilon_\textrm{eff}})^{-1}\label{eq:ResonanceFrequencies}
\end{equation}
where $c$ is the vacuum speed of light, $l$ the resonator's total length, and $\epsilon_\textrm{eff}$ the effective dielectric constant, which depends on the CPW geometry and the dielectric constants $\epsilon$ of the materials that are used, e.g. sapphire or TiO$_2$. The finite and temperature-dependent penetration depth for the superconducting film, which also affects the resonant frequency, we incorporate into the generic parameter $\epsilon_\textrm{eff}$. In spectroscopy applications, the frequency dependence of $\epsilon_\textrm{eff}$ is a key piece of information.

The microwave chips were mounted in brass boxes, and measurements of the complex transmission coefficient $\hat{S}_{21}$ through the feedlines were performed using a vector network analyzer (VNA) and a $^4$He cryostat with superconducting magnet and variable-temperature insert for temperatures $T$ down to 2\,K. 
The superconducting transition $T_\textrm{c}$ is around 8.6~K for device R$_\textrm{f}$ and around 7.1~K 
for device R$_\textrm{p}$.

For the microwave power-dependent measurements, amplifier and attenuator were used to reach higher power levels up to 17\,dBm.
Since our highest employed frequency is 20~GHz while the low-temperature superconducting energy gap of Nb is around 750~GHz,\cite{Pronin1998} we restrict our analysis using the assumption of frequency being much smaller than the energy gap, which might not rigorously hold for temperatures close to 
$T_\textrm{c}$.

From the $\hat{S}_{21}$ spectra, each resonance is fitted using the following function\cite{Thiemann_phd}:
%
\begin{equation}
\hat{S}_{21} = e^{\text{i}2\pi f\hat{\tau}} \left[ \frac{\hat{A}}{\left(f-f_\text{m}\right) +\text{i}\frac{f_\text{b,m}}{2}} +\hat{v}_3 + \hat{v}_4\left(f-f_\text{m}\right) \right]  \label{eq:FitFunctionResonances}
\end{equation}
%
Here $f_\textrm{m}$ is the resonance frequency, $f_\text{b,m}$ is the bandwidth where the generic index m includes designed and spurious resonances, $\hat{A}$ is a complex amplitude, $\hat{\tau}$ is a complex time constant, and the complex coefficients $\hat{v}_3$ and $\hat{v}_4$ model the background as first-order Taylor expansion. $Q_\textrm{m} = f_\text{m}/f_\text{b,m}$ is the experimentally observed loaded quality factor of the resonance. Real and imaginary parts of $\hat{S}_{21}(f)$ are fitted simultaneously. The discussion below will concentrate on $f_\textrm{m}$ and $Q_\textrm{m}$.

For a clear presentation of the numerous observed resonance modes, we use the following color coding in the figures below: 
for data obtained with R$_\textrm{f}$, the harmonic modes are plotted in shades of blue and black, with dashed and straight lines to distinguish adjacent modes.
For the very numerous modes analyzed for the R$_\textrm{p}$ device, the harmonic modes of the first resonator R$_\textrm{p1}$ are plotted in shades of blue, green, and yellow, and the modes of the second resonator R$_\textrm{p2}$ are plotted in shades of grey. Parasitic modes are plotted in shades of red for both R$_\textrm{f}$ and R$_\textrm{p}$ resonators.

\section{Results and Discussion}


\subsection{Spectra}

\begin{figure}
	\centering
	\includegraphics[width=\linewidth]{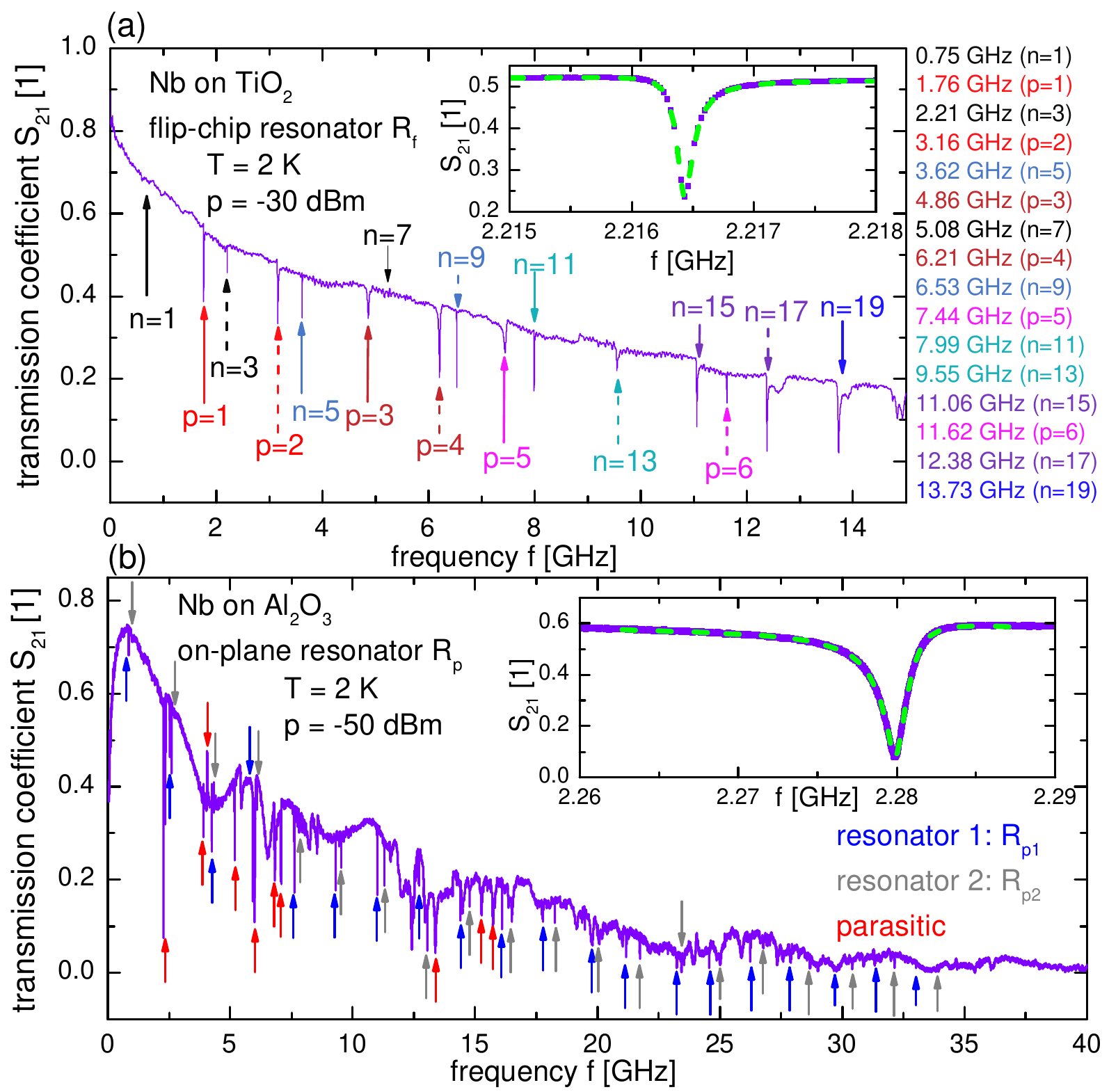}
	\caption{
	Transmission coefficient $|\hat{S}_{21}|$ of the (a) flip-chip setup R$_\textrm{f}$ and (b) on-plane setup R$_\textrm{p}$ measured at temperature $T=2\,$K. The insets show zoom-ins with fits to exemplary resonances. In (a), the labels of the resonances indicate the number $n$ of the harmonics (following Eq.\ \ref{eq:ResonanceFrequencies}) for the designated resonator modes whereas the number $p$ simply enumerates the parasitic modes that were analyzed. 
	}
	\label{fig:Spectra}
\end{figure}

In Fig.\ \ref{fig:Spectra} broadband spectra of the flip-chip setup R$_\textrm{f}$ and the on-plane setup R$_\textrm{p}$ are shown for $T=$ 2\,K. In both cases the background signal shows an overall decrease due to the transmission-line losses generally increasing with frequency for the CPW feedline and for the coaxial cables that connect the VNA and the cryogenic chip. 
Characteristic sharp minima in the spectra, indicated by arrows, arise for the designed harmonic modes as well as for the undesired parasitic resonances. In Fig.\ \ref{fig:Spectra}(a) the desired harmonic resonator modes are identified by their frequencies roughly equaling odd multiples of the fundamental frequency of 0.75~GHz, and the remaining resonances are labeled parasitic.
One reason why in Fig.\ \ref{fig:Spectra}(a) the frequencies of the harmonics are not exactly multiples of the fundamental frequency is the anisotropy of the TiO$_2$ combined with the varying contributions of the different crystallographic directions to the resonator response due to the standing wave pattern of the modes within the resonator. 
The assignment in Fig.\ \ref{fig:Spectra}(b) is complicated by the presence of two resonators but somewhat simplified by the less pronounced anisotropy of the sapphire substrate, therefore the odd multiples of the two fundamental resonances can be established straightforwardly where the remaining resonances are labeled parasitic again.
While all expected resonator harmonics are observed for the covered spectral range, the $n=7$ mode
of R$_\textrm{f}$ is very weak and thus this particular harmonic will not be considered below.


\subsection{Temperature Dependence} \label{sec:TemperatureDependence}


\begin{figure}
	\centering
	\includegraphics[width=\linewidth]{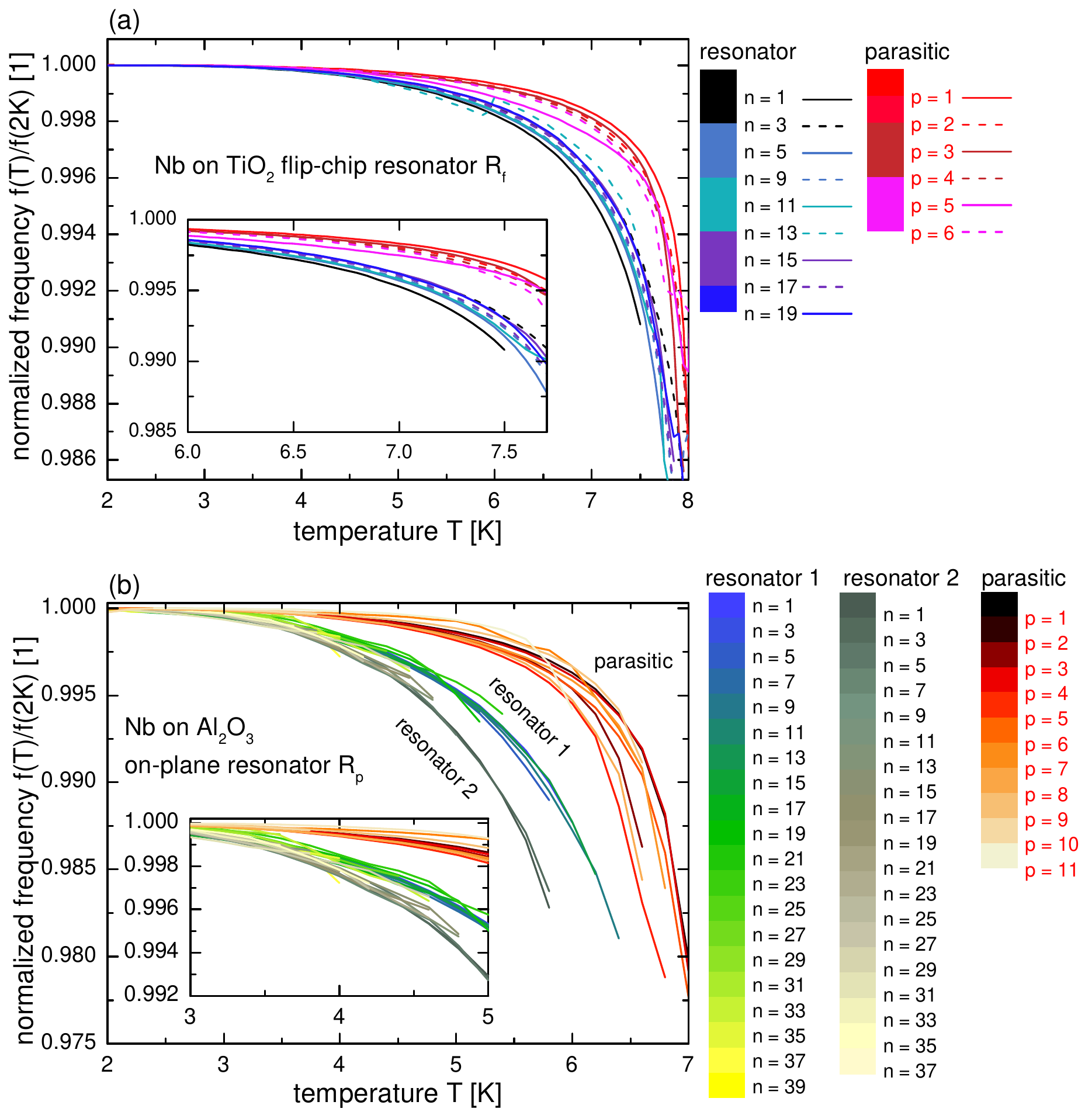}
	\caption{Resonance frequencies $f_\textrm{m}$ in dependence of the temperature $T$ for the R$_\textrm{f}$ case in (a) and R$_\textrm{p}$ case in (b). The resonance frequencies are normalized to the respective value at 2~K. The main panels show the complete temperature range, from 2~K up to the highest temperature where the modes were detected, and the insets show in more detail smaller temperature ranges.}
 \label{fig:TemperatureDependenceFrequency}
\end{figure}

The strong temperature dependence of superconducting properties can be used to distinguish designed and parasitic modes, as shown in Fig.\ \ref{fig:TemperatureDependenceFrequency}. For simpler comparison, the resonance frequencies are normalized to their respective values at 2~K, and while the main panels show the data for the full temperature range (from 2~K up to the highest temperature where the modes can still be properly distinguished from the background), the insets show in more detail the temperature ranges where the temperature-dependent evolution of the modes becomes evident.
One clearly sees that the data for the different modes form bundles of curves with similar behavior, and both for R$_\textrm{f}$ and for R$_\textrm{p}$ the parasitic modes have a weaker temperature dependence than the designated resonator modes. For a simple superconducting resonator based on a transmission line such as CPW, the transverse field distribution for all designated modes is equivalent, and therefore the temperature-dependent penetration depth of the superconductor will affect all resonator modes in the same fashion, via $\epsilon_\textrm{eff}$ in Eq.\ (\ref{eq:ResonanceFrequencies}),\cite{Hafner2014} and this is basically what one sees in Fig.\ \ref{fig:TemperatureDependenceFrequency}. Then it might come as a surprise that for the two CPW resonators of R$_\textrm{p}$  in Fig.\ \ref{fig:TemperatureDependenceFrequency}(b), which are fabricated within the same Nb layer and have the same lateral dimension of the CPW, the temperature evolution of the resonance frequencies is different with separating bundles of curves towards $T_\textrm{c}$. This can be explained if one assumes that the film quality of the Nb layer differs throughout different parts of the overall chip, and thus the \lq local $T_\textrm{c}$\rq{} might differ between resonators 1 and 2. Minute quality and thus $T_\textrm{c}$ variations within the Nb layer can also explain why the designated modes for each resonator, including the case in Fig.\ \ref{fig:TemperatureDependenceFrequency}(a), slightly differ in their temperature evolution.


\begin{figure}[t]
	\centering
	\includegraphics[width=\linewidth]{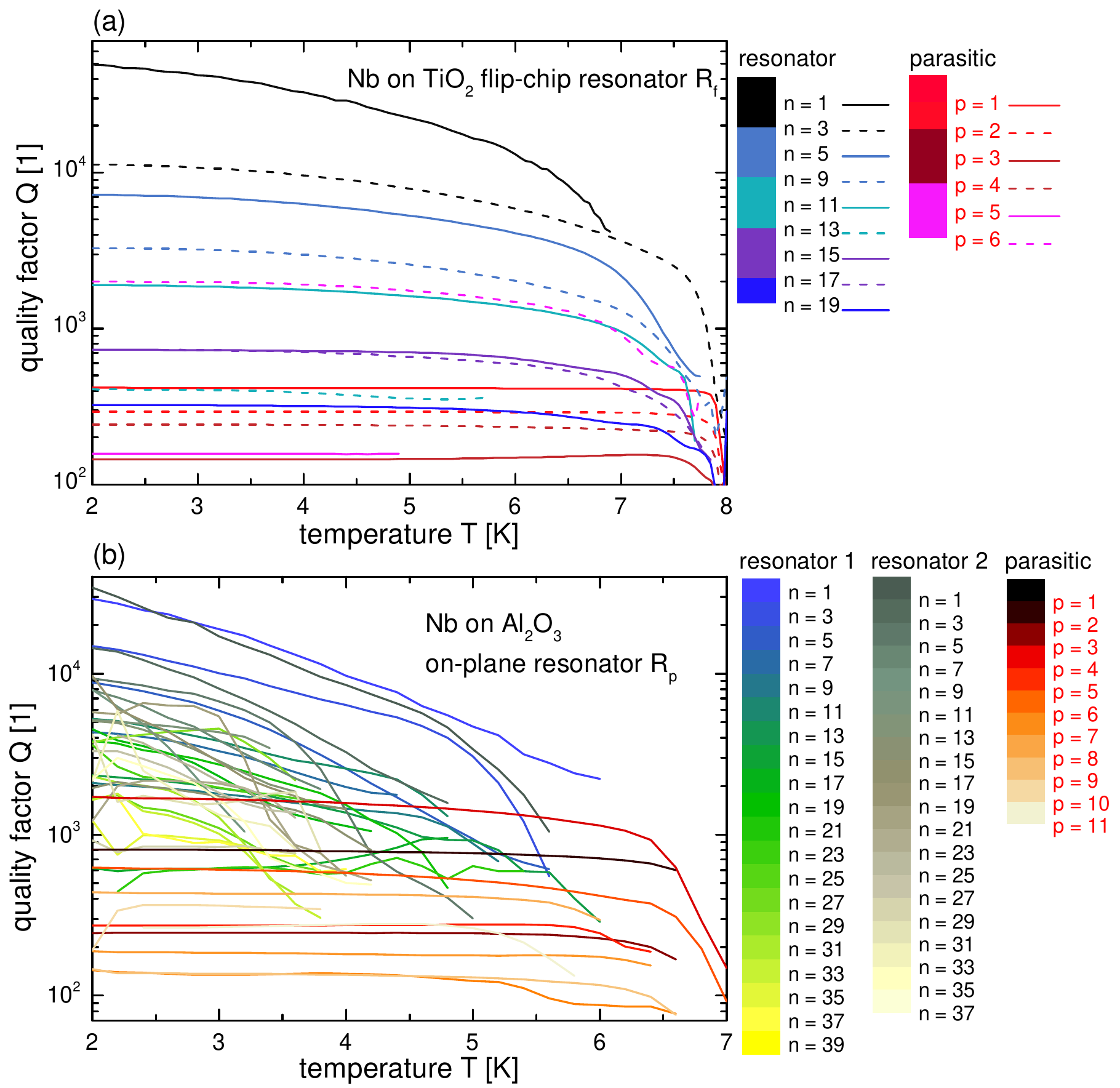}
	\caption{Measured quality factor $Q$ in dependence of the temperature $T$ for the R$_\textrm{f}$ case in (a) and R$_\textrm{p}$ case in (b) of the modes measured in Fig.\ \ref{fig:Spectra}.}
	\label{fig:TemperatureDependenceQ}
\end{figure}

\begin{figure}[t]
	\centering
	\includegraphics[width=\linewidth]{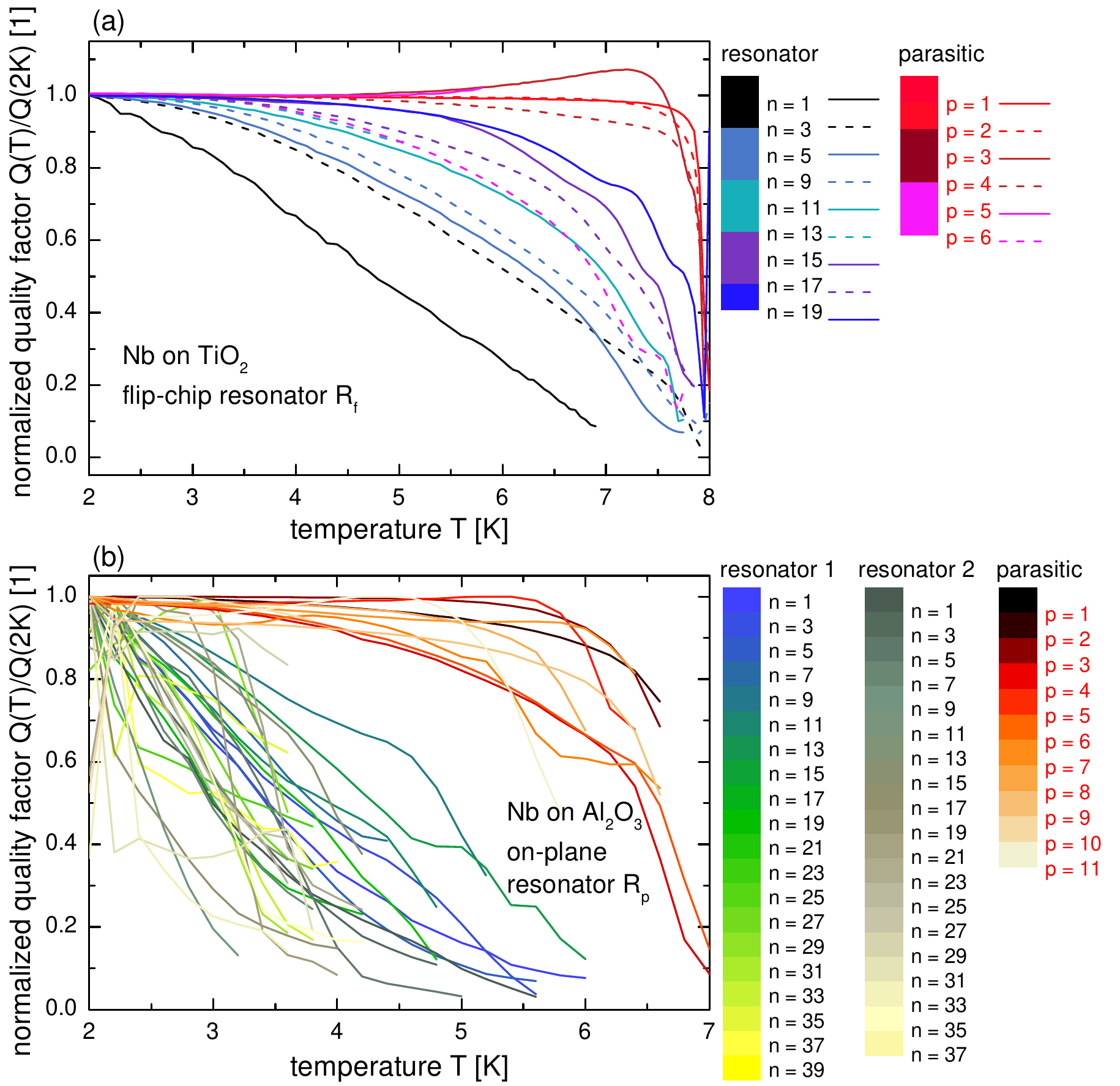}
	\caption{Measured quality factor $Q$ in dependence of the temperature $T$ for the R$_\textrm{f}$ case in (a) and R$_\textrm{p}$ case in (b) of the modes measured in Fig.\ \ref{fig:Spectra}. The quality factor is normalized to $Q(T=2\,\textrm{K})$}
	\label{fig:TemperatureDependenceQnormalized}
\end{figure}

Here we have assumed that the temperature dependence of $f_\textrm{n}$ is fully governed by the superconducting film. This assumption is justified in the present study because all other parameters that enter in Eq.\ \ref{eq:ResonanceFrequencies} can be assumed constant in this temperature range, e.g.\ the dielectric constants of TiO$_2$ or sapphire. \cite{Rausch2018, Sabinsky1962}

Also for the parasitic modes, which reside within the resonator chip and/or the housing box and thus can have as relevant further materials only metals, the superconducting film will have the strongest temperature dependence.
Indeed, the temperature dependence for parasitic modes in Fig.\ \ref{fig:TemperatureDependenceFrequency} is much less than the designated resonances. This means that for parasitic modes a much smaller fraction of the mode volume concerns the superconducting film. This matches the expectations for either undesired one-dimensional slotline modes of the CPW or three-dimensional modes that include the bulk of the substrate and/or the volume within the sample box.
The situation might be different if other strongly temperature-dependent materials are involved.\cite{Engl2019,Hering2007,Pompeo2007,Geiger2012}

The temperature dependence of the quality factor $Q$ is shown in Fig.~\ref{fig:TemperatureDependenceQ} for the different modes, and in Fig.~\ref{fig:TemperatureDependenceQnormalized} as normalized $Q(T)/Q(2\ \textrm{K})$.
Again one can clearly distinguish the desired resonator modes from the undesired parasitic ones, as they appear separated in Fig.\ \ref{fig:TemperatureDependenceQnormalized}: while the former decrease with increasing temperature already starting around 2\ K, the latter have much weaker temperature dependence and decrease substantially only close to $T_\textrm{c}$. But compared to the resonance frequencies in Fig.\ \ref{fig:TemperatureDependenceFrequency}, the $Q$ data do not assemble closely to bundles, and this has several reasons: firstly, the microwave losses of a superconducting resonator strongly depend on frequency, which is due to the characteristic low-frequency properties of the complex optical conductivity $\hat{\sigma}$ of superconductors,\cite{Pracht2013} and thus the absolute $Q$ of designated harmonics shown in Fig.\ \ref{fig:TemperatureDependenceQ} have a very strong frequency dependence in the low-temperature limit, roughly corresponding to $1/f$.\cite{Goeppl2008} In a similar fashion, the temperature evolution of $\hat{\sigma}$ also varies for different frequencies,\cite{Pracht2013,Steinberg2008} and thus no matching temperature dependence can be expected for the $Q$ of different resonant frequencies even when normalized (Fig.\ \ref{fig:TemperatureDependenceQnormalized}). Furthermore, there are various physical phenomena affecting $Q$. If there are separate loss mechanisms, one can assign a characteristic $Q$ to each of those, and the total, loaded $Q_\textrm{total}$ that we determine from the experiment is the inverse sum of the inverse respective $Q$s. For superconducting planar resonators, this might read as: 
\begin{equation}
  \frac{1}{Q_\textrm{total}} = \frac{1}{Q_\textrm{sc}} + \frac{1}{Q_\textrm{coupl}} + \frac{1}{Q_\textrm{diel}} + ... \label{eq:SumQs}
\end{equation}
where $Q_\textrm{sc}$ quantifies Ohmic losses in the superconductor, $Q_\textrm{coupl}$ coupling losses to the microwave readout, $Q_\textrm{diel}$ dielectric losses (in the substrate), and further contributions might consider radiation losses or losses in metallic components within the respective mode volume.
As discussed, $Q_\textrm{sc}$ strongly depends on frequency and temperature, and we have clear expectations based on the well-known $\hat{\sigma}(f,T)$ of conventional superconductors.\cite{Pracht2013,Steinberg2008} 
For the designated CPW modes, $Q_\textrm{diel}$ should be negligible here due to the choice of low-loss substrates, and $Q_\textrm{coupl}$, which is governed by geometrical parameters like $l_\textrm{c}$, for spectroscopy applications usually is designed to be rather high. In this case, $Q_\textrm{sc}$ represents the dominant loss channel and should obtain the strong frequency and temperature dependences discussed above. But if other mechanisms also contribute, e.g.\ quantified by a term $Q_\textrm{spur}$ of unclear origin that affect the spurious resonances and limit their $Q$ to values of order a few hundred, then the strong temperature dependence of possible $Q_\textrm{sc}$ contributions with absolute values above e.g.\ 1000 for temperatures well below $T_\textrm{c}$ will not affect much the $Q_\textrm{total}$ of the spurious modes, exactly as we see in Fig.\ \ref{fig:TemperatureDependenceQ}. Furthermore, the $Q$s of parasitic modes change little with temperature except close to $T_\textrm{c}$, in stark contrast to the designed CPW resonances.
These characteristics lead to the various intersecting curves in Fig.\ \ref{fig:TemperatureDependenceQ}, where $Q$s of designated modes clearly decrease with increasing temperature whereas $Q$s of the parasitic modes are almost constant.

\subsection{Magnetic Field Dependence}\label{sec:MagneticFieldDependence}

\begin{figure}
	\centering
	\includegraphics[width=\linewidth]{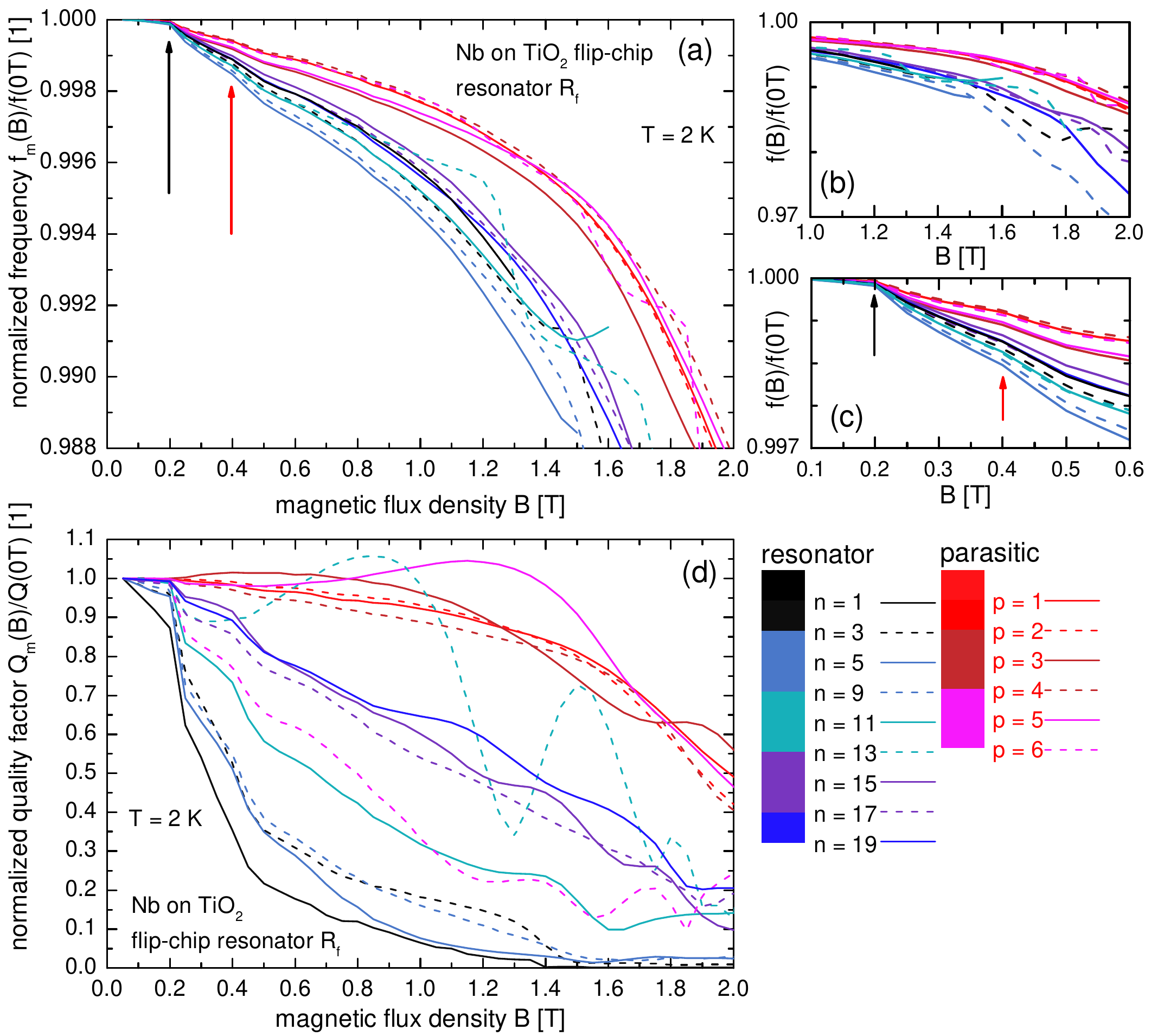}
	\caption{Resonance frequency $f_\text{m}$ in (a) and quality factor $Q_\textrm{m}$ in (d) normalized to the data at $B = 0$\,T for the TiO$_2$ flip-chip setup R$_\textrm{f}$ with the harmonic modes measured in Fig.\ \ref{fig:Spectra}, measured at $T=$ 2 ~K. (b) shows $f_\textrm{m}$ for $1\,\textrm{T}<B<2$\,T and (c) for $0.1\,\textrm{T}<B<0.6$\,T. The black arrow marks the starting point of decrease of $f_\textrm{m}$ at $B=0.2$\,T and the red arrow marks an abrupt change in decrease of $f_\textrm{m}$ at $B=0.4$\,T.}
	\label{fig:BfieldDependence}
\end{figure}

Applying an external static magnetic field $B$ has strong effects on superconductors, but basically leaves the other materials, such as the dielectric substrates, unaffected. For Nb, as type-II superconductor, the external magnetic field penetrates as quantized vortices for fields higher than the lower critical field $B_\textrm{c1}$ until superconductivity is fully suppressed (in the bulk) at the upper critical field $B_\textrm{c2}$. In our experiment, the external static magnetic field is applied roughly parallel to the Nb thin film of the resonator, and thus strong changes in the CPW performance are expected for fields of order 100~mT (in contrast to order 1~mT for perpendicular field).\cite{Bothner2012}
Due to this field arrangement and the strong dependence of superconducting properties on Nb material quality,\cite{Halbritter2005} it is difficult to quantitatively relate observed field-dependent effects to theoretical expectation. Still, the magnetic field dependence can help to assign resonator modes.

Fig.\ \ref{fig:BfieldDependence} shows the field dependence of $f_\textrm{m}$ and $Q_\textrm{m}$, both normalized to the respective zero-field values, at temperature $T = 2$~K for the R$_\textrm{f}$ device.
With increasing external static magnetic field and thus increasing vortex density, the microwave losses in the superconductor increase and thus result in a decreasing $f_\textrm{m}$, as clearly visible in Fig.~ \ref{fig:BfieldDependence}(a). 
Like for the temperature dependence in Fig.\ \ref{fig:TemperatureDependenceFrequency}, the designed and the parasitic modes assemble as well-separated bundles. The designated modes decrease more strongly with field due to the larger filling fraction of superconducting material within the mode volume compared to the parasitic modes. As indicated by arrows in Fig.~ \ref{fig:BfieldDependence}(a) and Fig.~\ref{fig:BfieldDependence}(c), two kinks in the data can be identified around 0.2~T and 0.4~T. From comparison with literature,\cite{Halbritter2005,Karasik1970,DasGupta1976} the first kink can be assigned to the first critical field $B_\textrm{c1}$, where vortices start to enter. The second kink indicates the second critical field $B_\textrm{c2}$, where superconductivity ceases in the bulk, while surface superconductivity continues for much higher static magnetic fields.

The field dependence of $Q_\textrm{m}$ in Fig.\ \ref{fig:BfieldDependence}(d) shows related behavior: again two kinks, for $B_\textrm{c1}$ and $B_\textrm{c2}$, can be identified around 0.2~T and 0.4~T. Above $B_\textrm{c1}$ the designed CPW modes exhibit strongly suppressed $Q_\textrm{m}$, while most of the parasitic modes are much less affected and hardly have any decrease in $Q_\textrm{m}$. Here one should keep in mind that the absolute zero-field $Q_\textrm{m}$ of the parasitics is already much lower than for the CPW modes.
The additional \lq oscillatory\rq{} field dependence of some of the modes (e.g.\ $n=11$, $n=13$, $p=3$, $p=5$, $p=6$) is due to overlap of the resonances in the microwave spectra with standing wave contributions of the background of the microwave spectra, which in these cases has not been fully covered by the fitting procedure and which changes as a function of field (and temperature) as the resonances move in frequency. This effect becomes more pronounced for broader resonances, and thus for higher fields and temperatures.

\subsection{Power Dependence}


\begin{figure}
	\centering
	\includegraphics[width=\linewidth]{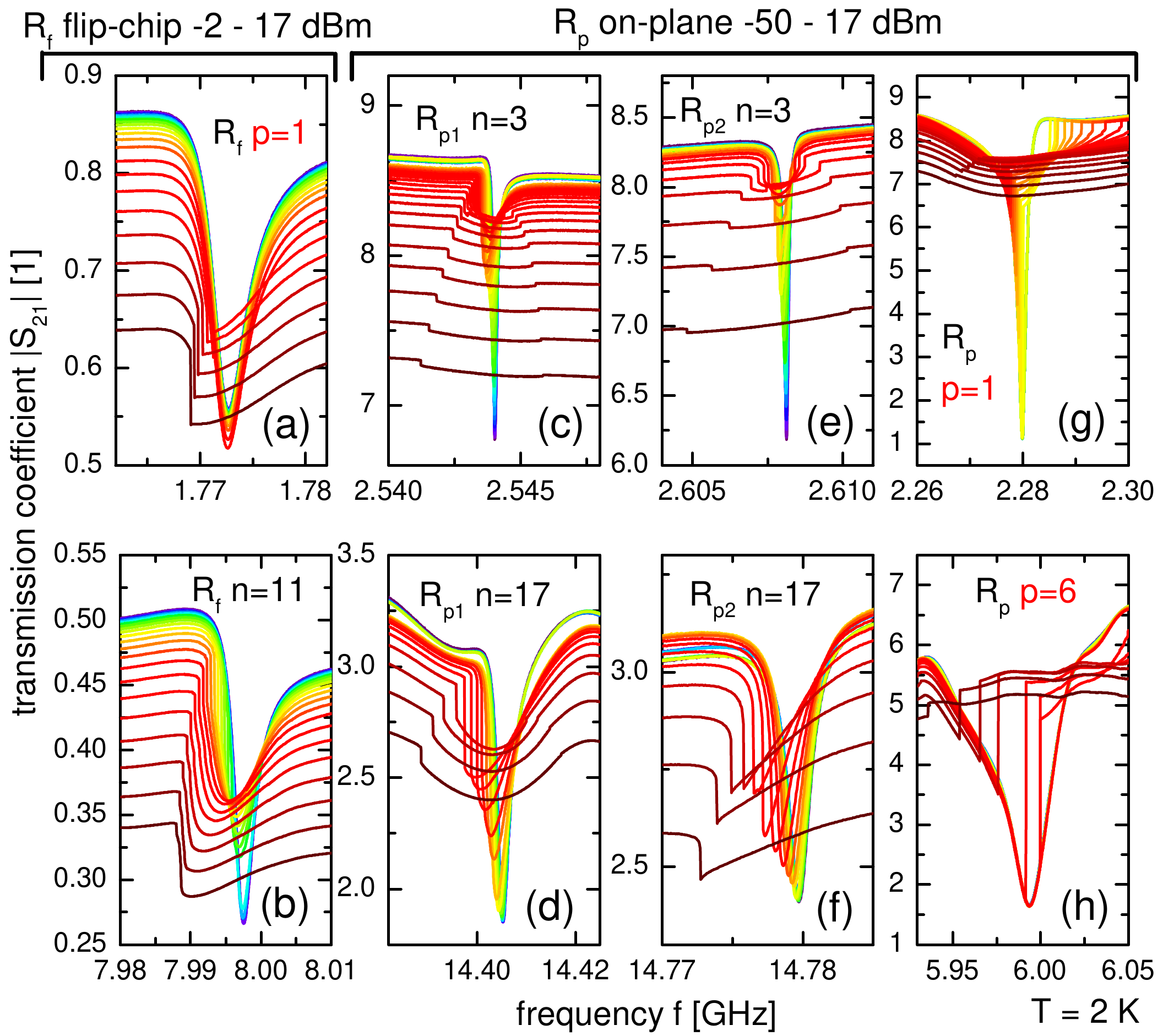}
	\caption{Measured spectra of transmission coefficient $|\hat{S}_{21}|$ for different harmonic and parasitic modes from the two setups $R_\textrm{f}$, with powers between -2~dBm (blue curve) and +17~dBm (brown curve) in steps of 1~dB, and $R_\textrm{p}$, with powers between -50~dBm (blue curve) and +17~dBm (brown curve) in steps of 3~dB. For greater powers $P$ anharmonicities arise which are characterized by sharp changes of $|\hat{S}_{21}|$. The decreased background of the spectra for high power is a result of the nonlinearity of the employed amplifier. 
	}
	\label{fig:PowerDependenceSpectra}
\end{figure}

\begin{figure}
	\centering
	\includegraphics[width=\linewidth]{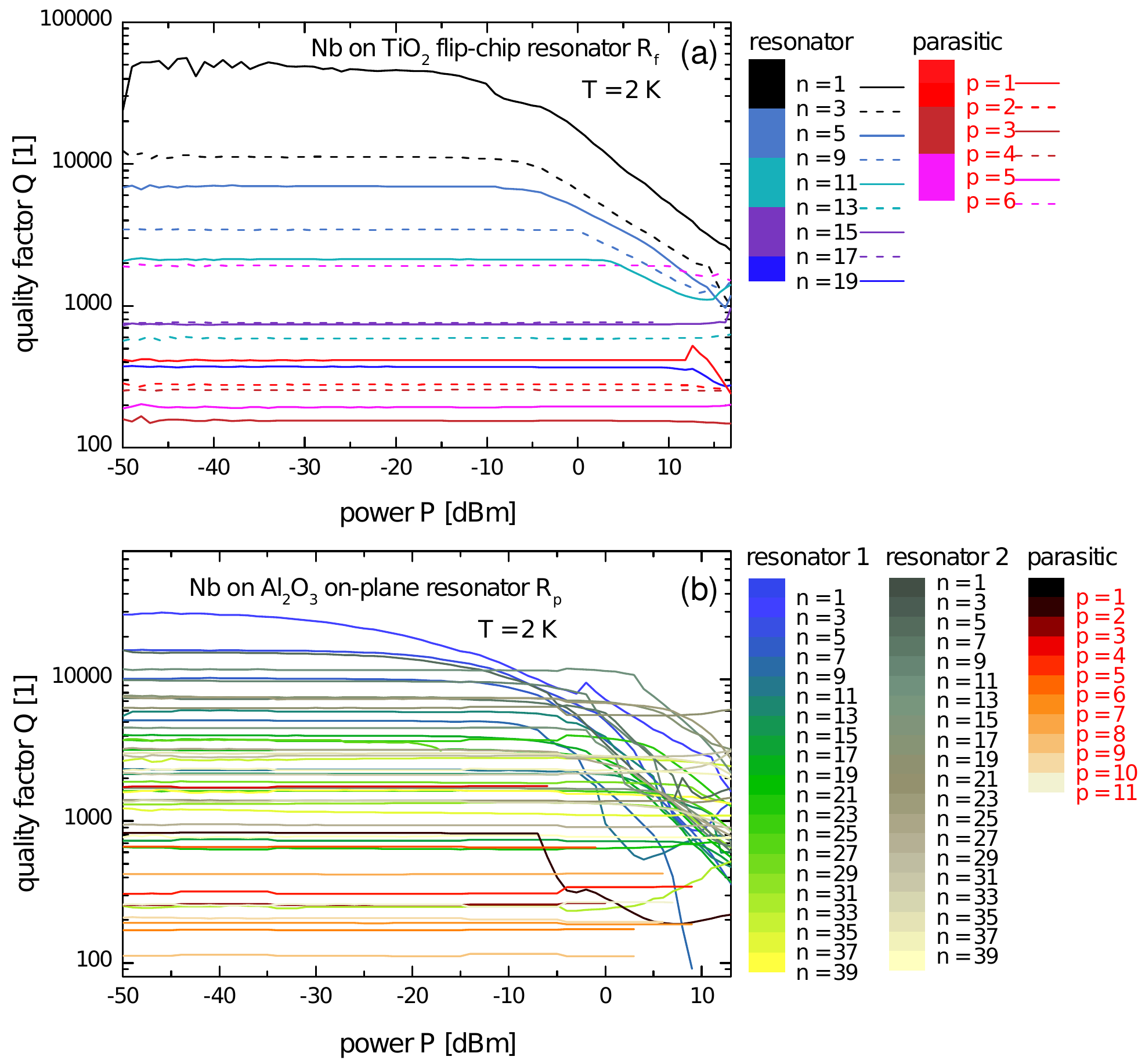}
	\caption{Quality factor $Q$ in dependence of the power $P$ for (a) the R$_\mathrm{f}$ case and (b) the R$_\mathrm{p}$ case, measured at $T =$ 2~K. }
	\label{fig:PowerDependenceQ}
\end{figure}

Another strategy to probe the nature of the different resonances concerns power dependence, i.e.\ studying the nonlinear behavior of the superconducting element. Here we focus on the behavior at temperature 2~K, i.e.\ much lower than $T_{\textrm{c}}$. 
With increasing power, basically three regimes are expected:\cite{Chin1992,Cohen2002,Abdo2006,deVisser2010} for low probing power, the resonator response is linear and $f_{\textrm{m}}$ and $Q_{\textrm{m}}$ are independent of power. 
For higher powers, the losses e.g\ due to thermally excited quasiparticles lead to a temperature increase, which in turn leads to a reduction of $f_{\textrm{m}}$ and $Q_{\textrm{m}}$, following the behavior discussed in Section \ref{sec:TemperatureDependence}. 
For even higher powers, the current density induced by the microwave field overcomes the critical current density of the superconductor at some position within the resonator. 
(The local microwave power depends strongly on the mode and its standing-wave pattern.) 
In this moment the superconductor turns normal at this position, and then the resonator properties change dramatically and exhibit certain characteristics of a metallic resonator, such as much lower $Q_{\textrm{m}}$ and $f_{\textrm{m}}$.

This generic behavior is indeed found in our data, as seen in Fig.\ \ref{fig:PowerDependenceSpectra} for various exemplary resonances of both devices R$_\text{f}$ and R$_\text{p}$. Considering the harmonic mode $n=11$ of R$_\text{f}$ in Fig.\ \ref{fig:PowerDependenceSpectra}(b), then one sees that for powers below +3 dBM (cyan curve) the resonance is basically unchanged for all powers. 
For the range from +3 dBM to +10 dBM (orange curve), the resonance becomes broader and weaker for increasing powers and shifts to lower frequencies, but the lineshape is still Lorentzian and can be properly fitted by Eq.\ \ref{eq:FitFunctionResonances} 
. For higher powers the situation changes drastically: e.g.\ for +17 dBM, there is an abrupt jump in the spectrum at 7.988 GHz 
. 
Above this frequency, the data follow a much broader resonance curve that is characteristic of the resonator being (at least partially) not superconducting any more but in the metallic state. In some cases, one can also clearly identify a second jump, back into the superconducting state, e.g.\ in Figs.\ \ref{fig:PowerDependenceSpectra}(c), (e), and (g) for resonator R$_\textrm{p}$.
Whenever jumps occur in the resonance spectra, it is not possible to determine unique values for $f_{\textrm{n}}$ and $Q_{\textrm{n}}$ for the full spectrum.
Comparable nonlinear behavior in planar superconducting resonators has been studied for various cases,\cite{Chin1992,Cohen2002,Abdo2006,deVisser2010} and different microscopic origins and theoretical descriptions have been discussed, but for our goal of just distinguishing different types of resonator modes we do not aim at a quantitative description of the nonlinear behavior.

If one fits the observed resonance spectra for all modes and powers to Eq.\ \ref{eq:FitFunctionResonances}, thus disregarding that the fit will not work well for spectra that include jumps, then one obtains power-dependent values of $f_{\textrm{m}}$ and $Q_{\textrm{m}}$. Here we focus on the behavior of $Q_{\textrm{m}}$ as shown in Fig.\ \ref{fig:PowerDependenceQ} for both devices, R$_\textrm{f}$ and R$_\textrm{p}$. For lowest tested powers, all modes have power-independent $Q_{\textrm{m}}$, thus indicating the linear regime. For higher powers, there are cases where $Q_{\textrm{m}}$ smoothly evolves into decreasing behavior and others where this decrease starts abruptly. The latter are those where jumps in the spectra set in at a critical power, and thus the spectra are not fitted properly any more.

When it comes to distinguishing regular resonator harmonics from parasitic modes, we find the general trend that the nonlinear behavior (decreasing $Q_{\textrm{m}}$) for resonator harmonics starts at lower powers than for the parasitic modes. This can be explained as follows: the nonlinear behavior sets in if the microwave-induced current density locally overcomes a certain threshold. 
For the designed modes, the microwave signal is directly induced into the CPW of the resonator and thus the largest current density is destined to flow in the center conductor with its rather small cross section. Even if we do not know the actual field distribution for the parasitic modes, we can assume that the current densities induced locally in the superconducting film are substantially smaller. 
This holds for \lq three-dimensional cavity modes\rq, where the microwave field is distributed throughout the comparably large volume of substrate(s) of the chip(s) as well as the sample box.
Also for the case of undesired slotline modes, the microwave electric field and thus the induced current density in the center conductor is smaller compared to the CPW mode, and thus a higher power has to be supplied to the overall device to induce strong nonlinearity for such a resonance.

\subsection{Dielectric Markers}\label{sec:DielectricMarkers}



\begin{figure}
	\centering
	\includegraphics[width=\linewidth]{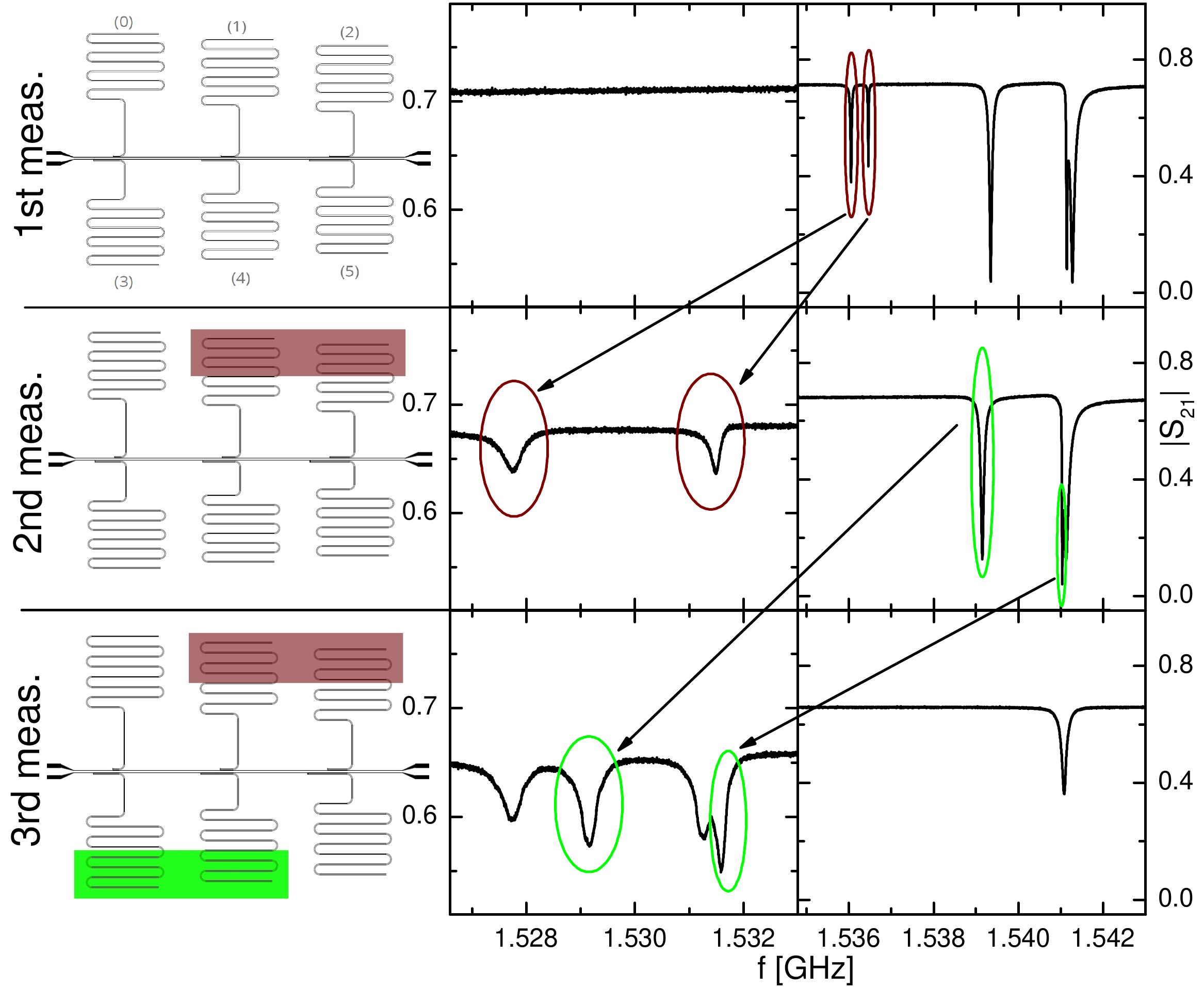}
	\caption{Identifying resonator modes by application of dielectric markers. The left column shows the design of the device R$_\text{p,diel}$ for three different cases: original device (upper row), device with dielectric markers attached to two of the resonators (middle row; area of dielectric markers shaded in red), device with dielectric markers attached to two further resonators (middle row; area of dielectric markers shaded in green).
The middle and right columns show the transmission spectra$|\hat{S}_\textrm{21}|$, measured at temperature 2~K, for the three different states of the device. Going from upper to middle row, two resonances shift due to the dielectric markers, and the same happens from middle to lower row.
}
	\label{fig:DielectricMarkers}
\end{figure}

If the in-situ strategies of the previous sections do not suffice to unambiguously assign observed resonances to specific modes of the device, one can minutely modify the resonator structure and observe which modes in the spectra then behave as expected. Considering Eq.~\ref{eq:ResonanceFrequencies}, one approach is changing $\epsilon_\textrm{eff}$ in a controlled fashion. 
For the designated CPW resonator modes, $\epsilon_\textrm{eff}$ includes a contribution due to the temperature-dependent penetration depth of the superconductor but to lowest order is the arithmetic mean of the dielectric functions of the dielectric substrate (here: TiO$_2$ or Al$_2$O$_3$) and vacuum/air/helium gas above the substrate. One can tune this by adding a small amount of another dielectric material on top of the CPW.\cite{Ebensperger2019,Wisbey2019}
Here we follow this strategy by using a conventional permanent marker pen.

In this case, we use a separate device, R$_\text{p,diel}$, that follows the overall design of the R$_\text{p}$ setup, but this new chip features six resonators as it can be seen in Fig.\ \ref{fig:DielectricMarkers}. All six resonators were designed to be at the same frequency, and thus the original spectrum of this device, shown in Fig.\ \ref{fig:DielectricMarkers}, features five resonances very close in frequency. The sixth resonator did not work properly. Here the task of mode assignment is extended such that one wants to identify which of the designated modes belongs to which resonator.
So two of the resonators were \lq marked\rq{} in a first step and two other resonators in a second step. 
The respective spectra with the fundamental modes around 1.54~GHz in Fig.\ \ref{fig:DielectricMarkers} clearly show how in each of these steps two of the resonances move to lower frequencies. 
These thus belong to the resonators where pigments of the marker pen were added, and therefore the respective modes can be assigned. 
This particular strategy resembles procedures that are being used in the field of KIDs, where resonator frequencies can be permanently adjusted e.g.\ by laser-trimming.\cite{Liu2017}
Our approach with a marker pen is less quantitatively predictable, but it can be implemented more easily and reversibly.

\subsection{ESR Markers}

\begin{figure}
	\centering
	\includegraphics[width=\linewidth]{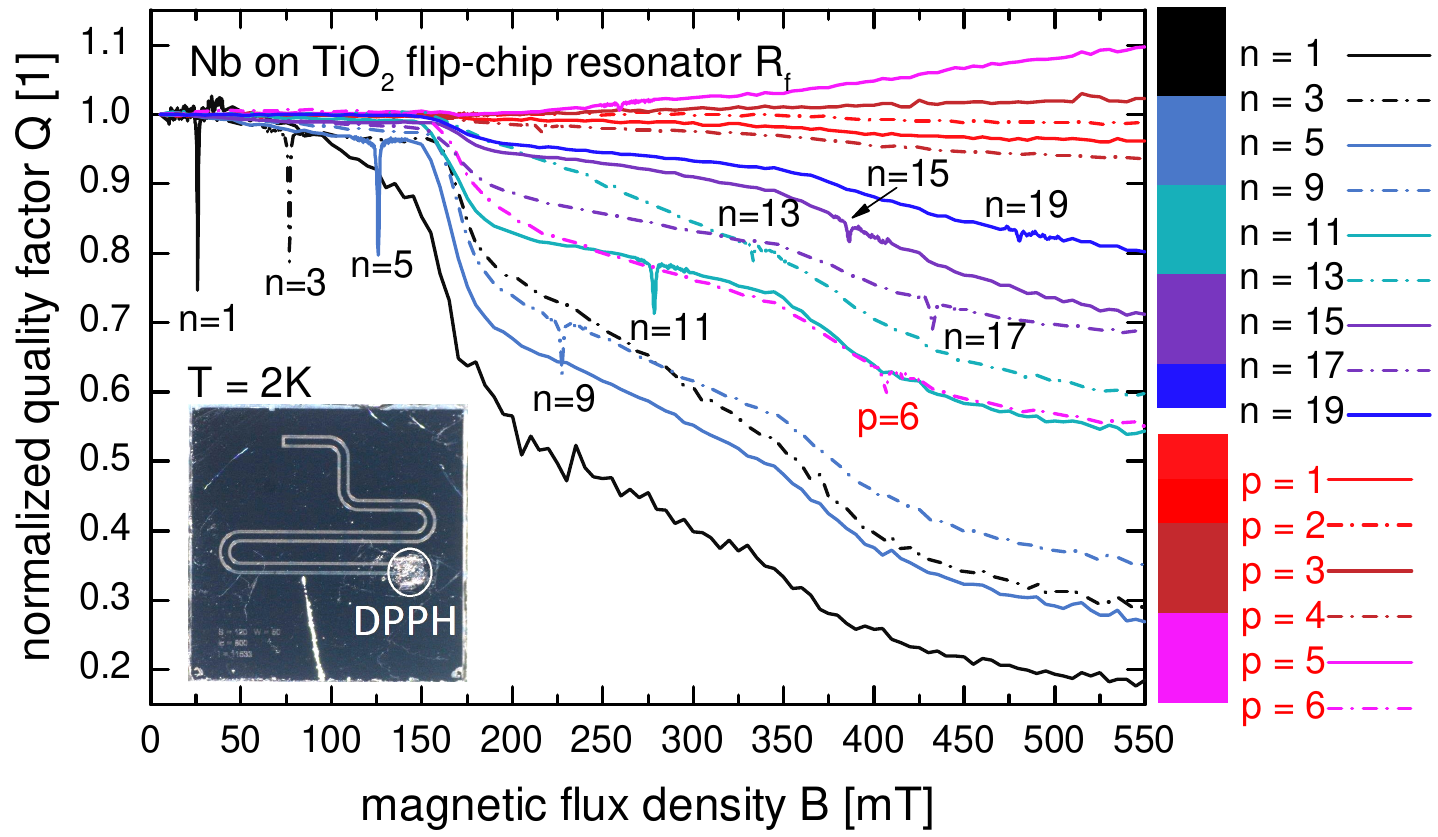}
	\caption{Normalized quality factor $Q$ of the harmonic and parasitic modes of the TiO$_2$ flip-chip resonator R$_\textrm{f}$ at $T = 2\,K$ in dependence of the static magnetic field $B$. Here $Q$ is normalized to its value at $B = 0\,$T. The DPPH sample is placed at the short-circuited end of the resonator (the end that is not adjacent to the feedline), as shown in the inset.
	}
  \label{fig:ESR1}
\end{figure}

\begin{figure}
	\centering
	\includegraphics[width=\linewidth]{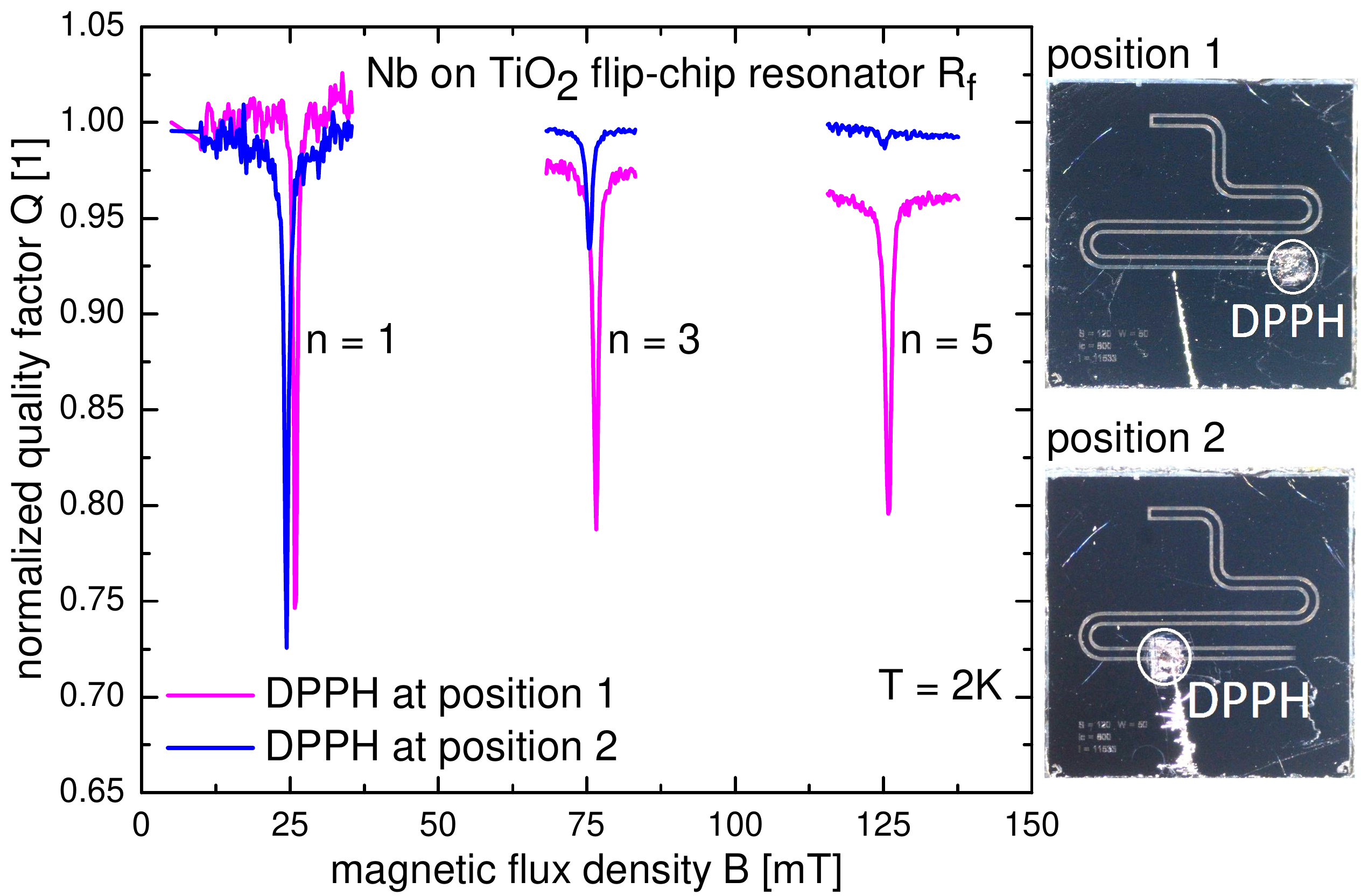}
	\caption{Normalized quality factor $Q$ for the $n = 1$, $n = 3$ and $n = 5$ modes for two separate measurements, where DPPH is applied to different locations, positions 1 and 2, as indicated on photographs on the right.}
	\label{fig:ESR2}
\end{figure}
%
%
%
The \lq dielectric marker\rq{} approach as presented above is hard to implement for the distant flip-chip design of R$_\textrm{f}$ because it would require removing the flip-chip from the sample box and later reattaching it, which for our way of mounting typically slightly changes the coupling between feedline and resonator chip, and thus basically all resonance frequencies, designed as well as parasitic, change somewhat.

Here, a different approach is possible that employs a \lq magnetic marker\rq{}. More specific, we use electron spin resonance (ESR) of the well-known paramagnet DPPH (1,1-diphenyl-2-picryl-hydrazyl) that is commonly used as reference material in ESR spectroscopy.\cite{Ghirri2015}
Magnetic effects are neglected in Eq.\ \ref{eq:ResonanceFrequencies} and in all discussions presented so far. This is justified because the frequency-dependent magnetic permeability for the relevant materials and settings of our experiments are very close to unity. This changes if the ESR condition holds:
\begin{equation}
	hf = g\mu_B B \label{eq:ESRCondition}
\end{equation}
with $f$ the frequency of a driving microwave magnetic field, $h$ Planck's constant, $g$ the Land\'{e} factor of the material (for DPPH $g\approx 2$), $\mu_B$ Bohr's magneton, and $B$ the external static magnetic field. 
For this combination of  $f$ and $B$, the microwave magnetic field component that is perpendicular to the external static magnetic field $B$ can induce transitions between the Zeeman-split energy levels of the material, and this means characteristic absorption of the microwave signal. For our case of resonance modes with almost fixed respective frequencies $f=f_\textrm{m}$ this means that if one sweeps the external static magnetic field $B$ and then fulfills Eq.\ \ref{eq:ESRCondition}, the microwave losses due to ESR will reduce the $Q_\textrm{m}$ of the mode at this particular $B$.

To take advantage of ESR for resonator mode identification, we apply a small amount of DPPH at a certain position of the resonator chip where we expect certain resonance modes to have strong microwave magnetic fields and thus strong ESR signal whereas other modes with weaker or absent microwave magnetic field at this position should exhibit weaker or absent ESR. Our resonator thus acts like an on-chip ESR spectrometer.\cite{Scheffler2013,Samkharadze2016,JavaheriRahim2016}

Fig.\ \ref{fig:ESR1} shows such an experiment, where the DPPH is deposited at \lq position 1\rq{} at the short-circuited end of the $\lambda/2$-type resonator of R$_\textrm{f}$: for all harmonics of the the CPW resonator, this position features a maximum of current and microwave magnetic field, and thus all harmonics should exhibit a clear ESR signal. This is indeed the case, see Fig.\ \ref{fig:ESR1}: the quality factors of the different modes as function of static external magnetic field show the overall evolution already known from Fig.\ \ref{fig:BfieldDependence}(a), but in addition there are pronounced, sharp minima. 
These occur at combinations of $f_\textrm{n}$ and $B$ according to the ESR condition, and thus their presence demonstrates that ESR can be used to encode information about certain resonance modes. As expected, all investigated CPW harmonics feature a clear ESR signal. 
In contrast, most of the observed ESR signals for parasitic modes at their respective $f_\textrm{m}$-$B$-combinations are weak. This is expected for three-dimensional cavity modes where the mode extends over a much larger volume than the one-dimensional CPW modes, and thus the microwave magnetic field at the position of the DPPH sample should be much weaker than for the CPW modes, leading to absence of observed ESR. 
One exception is the $p=6$ parasitic, which indeed features a pronounced ESR signal. This could mean that this mode is a slotline mode or a three-dimensional mode within the TiO$_2$ subtrate that \lq accidentally\rq{} features a substantial microwave magnetic field at the DPPH position. 

To further investigate the information that can be gained by ESR markers, we have performed another experiment with the DPPH deposited at a different position: we now choose \lq position 2\rq{} such that it should correspond to a microwave magnetic field node of the $n=5$ harmonic, and thus this mode should barely excite ESR. For the $n=3$ harmonic the microwave magnetic field should be substantially weaker compared to position~1 whereas for the fundamental $n=1$ harmonic there should only be a slight reduction and thus still strong ESR as before. In Fig.\ \ref{fig:ESR2} we show a close-up on the normalized quality factor $Q(B)$ for the $n=1,3,5$ modes for both discussed positions. As expected the ESR signal is almost completely suppressed for the $n=5$ mode, strongly reduced for the $n=3$ mode and slightly reduced for the $n=1$ mode when going with DPPH from position 1 to position 2. A small shift of the ESR signal to lower static magnetic fields $B$ can be observed which can be attributed to a small offset field of the superconducting magnet in the setup.

The ESR-marker technique is an elegant way to evaluate the microwave field strengths of different resonant modes at certain geometrical positions, and thus to verify the assignment of the modes as being dedicated resonator modes or parasitic. Compared to the dielectric markers of Section \ref{sec:DielectricMarkers} it has the advantage that the quality factor of any designated mode is substantially affected only near a single value of the external static magnetic field $B$, when Eq.~\ref{eq:ESRCondition} is met, and not affected for all other values of the external static magnetic field $B$ and thus possibly not interfering with other main experiments of interest.



\section{Summary}
This study examines differences between harmonic and parasitic modes of superconducting CPW resonators on a phenomenological level. Distinguishing these different types of modes can be important for the reliable interpretation of cryogenic microwave resonator data, and it can be particularly challenging if unconventional device geometries and/or materials with unknown microwave characteristics are involved.\cite{Engl2019} Therefore different mode assignment strategies have been presented, which can be grouped into those that analyze typically accessible microwave data of a given resonator structure and those that slightly modify the resonator structure to enable clearer mode assignment.

Tracking the resonance frequency $f_\textrm{m}$ of various modes as a function of temperature $T$ and external static magnetic field $B$ showed that designed harmonics and parasitic modes respectively form separate bundles in their decrease for increasing $T$ and $B$. This is due to the superconductor having a much larger filling fraction of the resonance mode volumes for the designed CPW modes compared to the parasitic modes. For the same reason, the quality factor $Q_\textrm{m}$ of the resonator harmonics exhibits stronger temperature and magnetic-field dependence compared to parasitic modes. Also in the power dependence much stronger nonlinear effects are observed for the designed harmonic modes compared to the parasitic ones. If such data sets are not sufficient to unambiguously assign the modes, one can add small amounts of dielectric and/or ESR markers to selectively tune some of the modes, and then check for the expected changes in the microwave response.
While any of the presented techniques might be sufficient for mode assignment, we found that in the more challenging cases the combination of several of them is most convincing.

\section*{Acknowledgments}
We thank D. Bothner for helpful discussions and Deutsche Forschungsgemeinschaft (DFG) for financial support.


\end{document}